\documentclass{iaus}
\usepackage{graphicx,url,bm}
\newcommand{\EQ}{\begin{equation}}
\newcommand{\EN}{\end{equation}}
\newcommand{\EQA}{\begin{eqnarray}}
\newcommand{\ENA}{\end{eqnarray}}

\newcommand{\Fig}[1]{Fig.~\ref{#1}}

{}
{}

{}
{}
{}
{}
{}
{}
{}
{}
{}
{}
{}

\newcommand{\meanU}{\overline{U}}

\newcommand{\pphi}{\hat{\bm{\phi}}}

\newcommand{\zzz}{\mbox{\boldmath $z$} {}}

\newcommand{\UU}{\mbox{\boldmath $U$} {}}

\newcommand{\FF}{\mbox{\boldmath $F$} {}}

\newcommand{\grav}{\mbox{\boldmath $g$} {}}
\newcommand{\nab}{\mbox{\boldmath $\nabla$} {}}
\newcommand{\OO}{\mbox{\boldmath $\Omega$} {}}

\newcommand{\SSSS}{\mbox{\boldmath ${\sf S}$} {}}

\newcommand{\DD}{{\rm D} {}}

\newcommand{\dd}{{\rm d} {}}

\def\half{{\textstyle{1\over2}}}

\def\onethird{{\textstyle{1\over3}}}

\newcommand{\nHz}{\,{\rm nHz}}

\newcommand{\mpers}{\,{\rm m/s}}

\newcommand{\Mm}{\,{\rm Mm}}

\newcommand{\yapj}[3]{ #1, {ApJ,} {#2}, #3}

\newcommand{\yapjl}[3]{ #1, {ApJ,} {#2}, #3}

\newcommand{\yan}[3]{ #1, {Astron.\ Nachr.,} {#2}, #3}

\newcommand{\yana}[3]{ #1, {A\&A,} {#2}, #3}

\newcommand{\ygafd}[3]{ #1, {Geophys.\ Astrophys.\ Fluid Dyn.,} {#2}, #3}

\newcommand{\yjfm}[3]{ #1, {J.\ Fluid Mech.,} {#2}, #3}

\newcommand{\yaraa}[3]{ #1, {ARA\&A,} {#2}, #3}

\newcommand{\ynat}[3]{ #1, {Nature,} {#2}, #3}
\newcommand{\ysci}[3]{ #1, {Science,} {#2}, #3}
\newcommand{\ysph}[3]{ #1, {Solar Phys.,} {#2}, #3}

\newcommand{\yjour}[4]{ #1, {#2}, {#3}, #4}

\newcommand{\ybook}[3]{ #1, {#2} (#3)}

\title[Near-surface shear layer] %% give here short title %%
{Near-surface shear layer dynamics}

\author[A. Brandenburg]   %% give here short author list %%
{Axel Brandenburg}

\affiliation{Nordita, Blegdamsvej 17, DK-2100 Copenhagen \O, Denmark, and
\break AlbaNova University Center, SE - 106 91 Stockholm, Sweden
\break email: brandenb@nordita.dk}

\pubyear{2007}
\volume{xxx}  %% insert here IAU Symposium No.
\pagerange{1--12}
\date{?? and in revised form ??}
\jname{Proceedings Title IAU Symposium}
\editors{A.C. Editor, B.D. Editor \& C.E. Editor, eds.}
\begin{document}

\maketitle

\begin{abstract}
The outer surface layers of the sun show a clear deceleration
at low latitudes. This is generally thought to be the result of a
strong dominance of vertical turbulent motions associated with
strong downdrafts. This strong negative radial shear should not only
contribute to amplifying the toroidal field locally and to expelling
magnetic helicity, but it may also be responsible for producing a strong
prograde pattern speed in the supergranulation layer. Using simulations
of rotating stratified convection in cartesian boxes located at low
latitudes around the equator it is shown that in the surface layers
patterns move in the prograde direction on top of a retrograde mean
background flow. These patterns may also be associated with magnetic
tracers and even sunspot proper motions that are known to be prograde
relative to the much slower surface plasma.
\end{abstract}

\firstsection % if your document starts with a section,
\section{Introduction}

The existence of the near-surface shear layer has become strikingly
clear only in recent years.
Helioseismology has always predicted rotation frequencies near the surface
that were in the range of about 460--470 nHz, while the photospheric
value determined by Doppler measurements was always around 455 nHz.
The reason for this apparent mismatch became clear when the spatial
resolution became high enough so that a decline of the angular
velocity toward the surface could actually be measured in detail.

A negative $\Omega$ gradient was always natural to expect,
because angular momentum conservation associated with the vertical
mixing of fluid elements would lead to $\Omega\sim1/r^2$
(Wasiutynski 1946, Kippenhahn 1963).
Prior to the days of helioseismology one expected that a negative
$\partial\Omega/\partial r$ gradient would extend through the entire
convection zone.
Another piece of supporting evidence came from the decrease in the
sunspot proper motions: $473\nHz$ for the youngest sunspots (0.5--1.5 days
old), $462\nHz$ for the oldest sunspots (3 months old), and only
$452\nHz$ for the angular velocity of the surface plasma as determined
from the Doppler shift.
With the beginning of the 1980ties one expected sunspots to be anchored
deep down at the bottom of the convection zone, because magnetic buoyancy
would be too strong near the upper parts of the convection zone
(Spiegel \& Weiss 1980).
As sunspots become older, one thought they might be rooted less deep,
reflecting therefore the angular velocity in the upper layers, thus
suggesting a negative $\partial\Omega/\partial r$ gradient.

In the mean time, with the advent of helioseismology, it became clear
that throughout most of the convection zone the angular velocity is
nearly independent of radius while at the bottom of the convection
zone there is a pronounced negative radial gradient.
With this the earlier explanation of the enhanced sunspot proper
motion was no longer applicable.

Over the past few years two related features of the near-surface dynamics
of the sun have received increased attention: the presence of a marked
shear layer near the surface and prograde pattern speeds in the
supergranulation and in magnetic tracers.
Already the early helioseismic inversions by Duvall et al.\ (1984)
suggested angular velocities in the upper parts of the sun that where
significantly faster than the photospheric plasma as determined by
Doppler shifts.
On the other hand, sunspots have long been known to spin about 5\%
faster than the photospheric plasma (Gilman \& Foukal 1979,
Golub et al.\ 1981).
With recent helioseismic inversions it became quite clear that there is
indeed a sharp shear layer near the surface where the local rotation rate
decreases from a value of more than $470\nHz$ at a depth of about $35\Mm$
to $455\nHz$ at the surface (e.g., Howe et al.\ 2000, Thompson et al.\ 2003).
Furthermore, using Doppler velocity images of the sun, Gizon et al.\ (2003)
found phase speeds of about $65\mpers$, which may also explain why the
supergranulation rotates faster than the surface plasma.

Theoretically, a negative radial gradient of the angular velocity may be
understood as a consequence of a local dominance of vertical convective
motions over radial ones (R\"udiger 1980, 1989).
The theoretically predicted connection between velocity anisotropy and
mean flow shear has also been tested in various simulations
(Pulkkinen et al.\ 1993, Chan 2001, K\"apyl\"a et al.\ 2004).
Furthermore, new detailed mean field models now begin to reproduce also
the near surface shear layer correctly (Kitchatinov \& R\"udiger 2005).
More surprising is the suggestion that supergranulation and even sunspots
and other magnetic tracers could essentially be a wave pattern with a
prograde pattern speed.
Linear theory confirms that this would indeed be possible (Busse 2004),
although the pattern speed would still be too small to explain the observed
pattern speed (Green \& Kosovichev 2006).
Alternative interpretations have been offered by Hathaway et al.\ (2006),
who proposed that the enhanced pattern speeds are simply the result
of a projection effect, and by Rast et al.\ (2004), who argue that a
spuriously wave-like spectrum is obtained when the image tracking rate
falls between the actual mesogranular and supergranular rotation rates.
In the present paper we focus on the possibility that there are true
traveling wave patterns and investigate their speed in the nonlinear
regime.

\section{Simulations}

We consider here a recent model considered by Brandenburg \& Kosovichev
(unpublished) where they modeled
convection in a slab of size $L_x\!\times L_y\!\times L_z$ with
periodic boundary conditions in the horizontal directions ($x$ and $y$)
and free-slip boundary conditions in the vertical direction ($z$).
The correspondence with spherical polar coordinates is
$(r,\theta,\phi)\leftrightarrow(z,x,y)$, so $x$ points south and $y$ east.
The equations for a compressible fully ionized perfect gas are solved
in the form
\EQ
{\DD\ln\rho\over\DD t}=-\nab\cdot\UU,
\EN
\EQ
{\DD\UU\over\DD t}=-c_{\rm s}^2(\nab\ln\rho+\nab s/c_p)
-2\OO\times\UU+\grav+\FF_{\rm visc},
\label{dudt}
\EN
\EQ
T{\DD s\over\DD t}=
2\nu\SSSS^2+{1\over\rho}\nab\cdot K\nab T-{\cal L}_{\rm surf},
\EN
where $\UU$ is the velocity, $\rho$ the density, $s$ the specific entropy,
$\OO=(\sin\theta,0,\cos\theta)\Omega$ is the angular velocity,
so $-\theta$ corresponds to the polar angle,
$\grav=(0,0,-g)$ is the gravitational acceleration,
$\FF_{\rm visc}=\rho^{-1}\nab\cdot(2\nu\rho\SSSS)$ is the viscous force, where
${\sf S}_{ij}=\frac{1}{2}(U_{i,j}+U_{j,i})-\frac{1}{3}\delta_{ij}\nab\cdot\UU$
is the traceless rate of strain tensor,
$\nu$ is the kinematic viscosity, $\chi$ is the thermal diffusivity,
and ${\cal L}_{\rm surf}$ is a net cooling/heating applied in the
surface layer with
\EQ
{\cal L}_{\rm surf}=f(z){c_{\rm s}^2-c_{\rm s0}^2\over\tau_{\rm cool}},
\EN
where $f(z)=\half\{1+\tanh[(z-d)/w]\}$ is a profile function of width $w=0.05$,
and $\tau_{\rm cool}=\sqrt{d/g}$ has been chosen for the relaxation time.

A convectively unstable layer of depth $d$ in $0<z<d$ is confined between
stable layers at top and bottom (at $z/d=1.15$ and $-0.85$, respectively).
This is achieved by adopting a $z$ dependent profile for the radiative
conductivity $K$.
A convenient nondimensional measure for the conductivity is
(Brandenburg et al.\ 1996; 2005)
\EQ
m+1={gK\over(c_p-c_v)F_{\rm tot}},
\label{m+1}
\EN
where $g$ is the gravitational acceleration, $c_p$ and $c_v$ are the
specific heats at constant pressure and constant volume, respectively,
and $F_{\rm tot}$ is the total energy flux.
In the following we choose $m+1=0.5$ in the unstable layer (i.e.\ $m=-0.5$),
and $m=6$ in the stable layer beneath.
In the upper stable layer a constant temperature profile with
$c_{\rm s}\approx0.1$ is enforced by adopting a heating/cooling term,
${\cal L}_{\rm surf}$, as defined above.
By taking $g=1$, we may define units for time and velocity as
$\sqrt{d/g}$ and $\sqrt{gd}$, respectively.
Nevertheless, for clarity we often keep the units explicitly.
We choose $L_x=L_y=4d$ and $L_z=2d$ with a resolution of
$512\times512\times256$ meshpoints.

\begin{figure}[t!]\begin{center}
\includegraphics[width=.7\columnwidth]{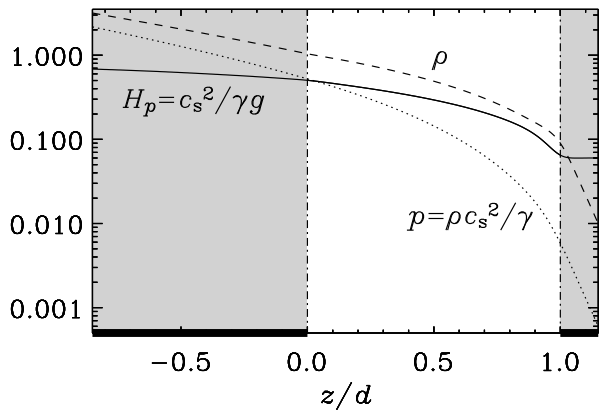}
\end{center}\caption[]{
Mean stratification of density, pressure, and pressure scale height.
The shaded areas mark the stably stratified layers.
}\label{pphp_256_512a_Om02}\end{figure}

We adopt a perfect gas where $\rho$ and
$s$ are related to temperature $T$ and sound speed via the relations
\begin{equation}
(\gamma-1)c_pT=c_{\rm s}^2=gd\exp[(\gamma-1)\ln(\rho/\rho_0)+s/c_v],
\label{eos}
\end{equation}
where $\gamma\equiv c_p/c_v=5/3$, and $\rho_0$ is a reference density
that is chosen to agree with the initial density at the bottom of the
unstable layer ($z=0$).
This essentially fixes the density scale such that the mass in
the system is close to unity.
The initial condition is constructed from a simple mixing length model
that is calibrated to the simulations (Brandenburg et al.\ 2005).
With this initial condition the density at $z=0$ remains close to its
initial value.

As boundary conditions we assume stress-free conditions at top and bottom,
and an imposed flux at the bottom.
At the bottom a constant total energy flux, $F_{\rm tot}$, is
given by specifying the non-dimensional number
${\cal F}=F_{\rm tot}/[\rho_0(gd)^{3/2}]$.
For the adopted set of parameters ${\cal F}$ is related to the Rayleigh
number via $\mbox{Ra}\approx4.7/{\cal F}^2$ (see Brandenburg et al.\ 2005
for details).
For most of the runs presented below we take ${\cal F}=5\times10^{-4}$,
corresponding to $\mbox{Ra}=1.9\times10^7$.
The angular velocity, $\Omega$, is chosen to be $\Omega=0.2\sqrt{g/d}$.
This would be a serious exaggeration if compared with solar values,
but the dynamically relevant parameter is the inverse Rossby number,
$\mbox{Ro}^{-1}\!=2\Omega H_p/u_{\rm rms}$, where $H_p=c_{\rm s}^2/(\gamma g)$
is the pressure scale height.
The value of $\mbox{Ro}^{-1}$
depends on $z$ and varies between $\sim0.08$ at the top ($z=d$)
to about 6 at the bottom of the unstable layer ($z=0$), which compares
favorably with the sun below some $30\Mm$ depth.

\section{Results}

\begin{figure}[t!]\begin{center}
\includegraphics[width=.7\columnwidth]{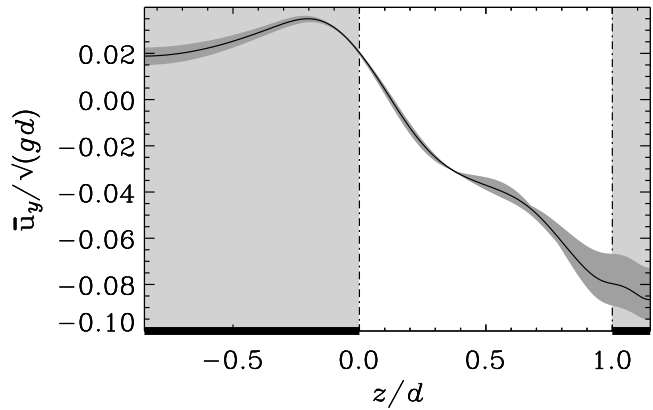}
\end{center}\caption[]{
Mean azimuthal flow, $\meanU_y(z)$.
The light shaded areas indicate the stably stratified zones, while
the dark gray envelope around the curve indicate the error
of $\meanU_y(z)$, estimated from the maximum deviation found
from the 3 averages over only 1/3 of the time interval.
$0^\circ$ latitude.
$512\times512\times256$ resolution, $4\times4\times1$ aspect ratio.
}\label{ppuymzm_256_512a_Om02}\end{figure}

We have performed runs for different values of $\theta$ corresponding
to the latitudes between $0^\circ$ and $30^\circ$.
The longest run with the best statistics is for $0^\circ$ latitude
($-\theta=90^\circ$).
In \Fig{pphp_256_512a_Om02} we plot the average stratification
of specific entropy, temperature, and pressure.
Next, we consider the mean toroidal flow for simulations with different
values of $\theta$.
The mean azimuthal flow speed,
\EQ
\meanU_y\equiv{1\over L_xL_y}\int U_y\;\dd x\,\dd y
\EN
is negative in the upper layers; see \Fig{ppuymzm_256_512a_Om02}.
A similar negative shear profile has also been found in recent simulations
of deep solar-like convection by Stein \& Nordlund (private communication).
Note that in our runs the shear is
$\dd\meanU_y/\dd z\approx-0.1\sqrt{g/d}=-0.5\Omega$,
which is rather strong.
Nevertheless, both magnitude and the negative sign are comparable to the
Reynolds stress, $\overline{u_y'u_z'}$, where primes denote deviations
from the horizontal mean.
Indeed, the toroidal component of the steady-state momentum
equation, $\overline{u_yu_z}$, turns out to be similar to
$\nu_{\rm t}{\partial\meanU_y\over\partial z}$,
if $\nu_{\rm t}\approx30\nu$ is assumed.
The ratio $\nu_{\rm t}/\nu$ is a measure of the length of the inertial
range of the turbulence and hence of the fluid Reynolds number based
on the flow properties at the energy carrying scale.
The value of $\nu_{\rm t}$ is compatible with the standard estimate
$\nu_{\rm t}=\onethird u_{\rm rms}\ell$, where the mixing length $\ell$
can be estimated by the pressure scale height $H_p$.

\begin{figure}[t!]\begin{center}
\includegraphics[width=.7\columnwidth]{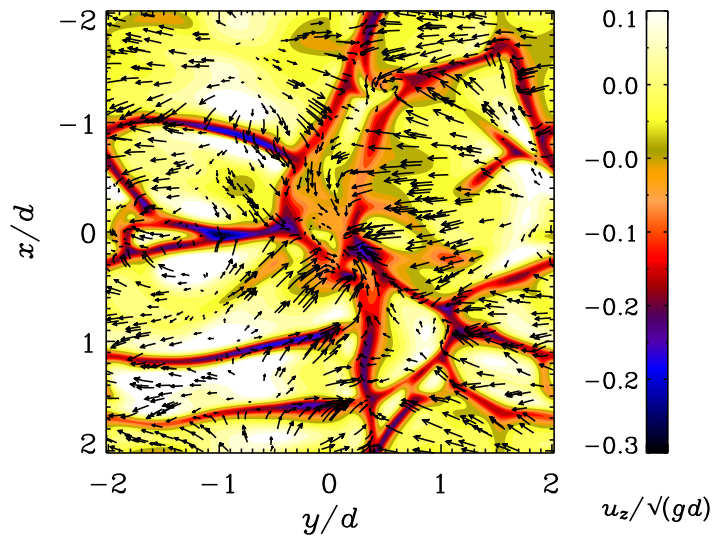}
\end{center}\caption[]{
Horizontal slice of a snapshot at an arbitrarily chosen time
$t=254/\sqrt{g/d}$ at height $z=0.9d$.
Note the granular pattern elongated in the azimuthal direction,
with dark lanes corresponding to downward motion ($u_z<0$).
$0^\circ$ latitude.
$512\times512\times256$ resolution, $4\times4\times1$ aspect ratio.
}\label{pslice_256_512a_Om02}\end{figure}

\begin{figure}[t!]\begin{center}
\includegraphics[width=.7\columnwidth]{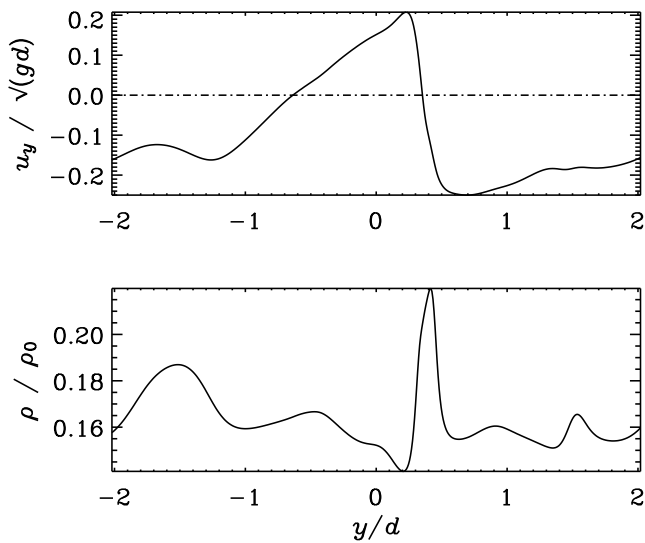}
\end{center}\caption[]{
Azimuthal profile of the azimuthal velocity, $u_y(y)$, and density,
$\rho(y)$, through $x=-2d$ and $z=0.9d$ for $t=254/\sqrt{g/d}$;
see upper and lower panels, respectively.
Note the shock-like profile with prograde velocity and a weak
density enhancement riding on a retrograde moving background.
This figure is for the same snapshot as \Fig{pslice_256_512a_Om02}.
}\label{pslice_lnrho_256_512a_Om02}\end{figure}

\begin{figure}[t!]\begin{center}
\includegraphics[width=.7\columnwidth]{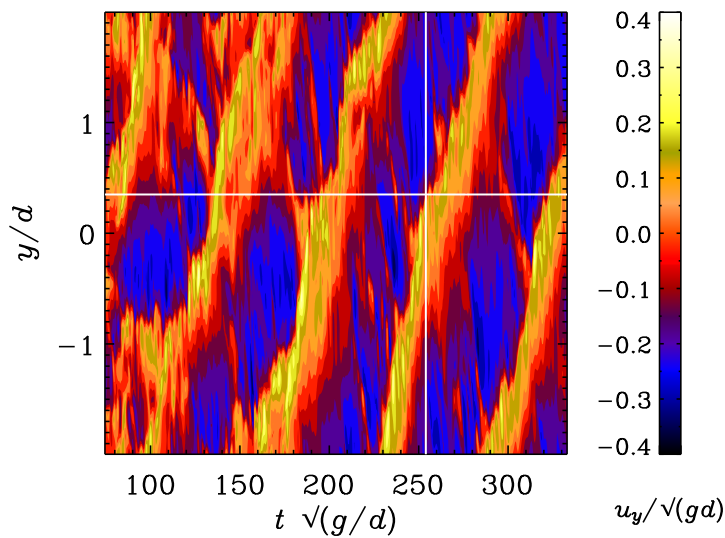}
\end{center}\caption[]{
Mean azimuthal flow speed at $x=-2d$ and $z=0.9d$.
The pattern speed inferred from this graph is $+0.064\sqrt{gd}$.
The azimuthal flow speed inside the light stripes exceeds the
pattern speed and is $+0.15\sqrt{gd}$, while the flow
speed between the stripes is $-0.25\sqrt{gd}$, such that
the average speed is $-0.1\sqrt{gd}$.
This figure is for the same run as in \Fig{pslice_256_512a_Om02}.
The white lines denote the time $t=254/\sqrt{g/d}$ and `shock'
position $y=0.35d$, as seen in \Fig{pslice_lnrho_256_512a_Om02}.
}\label{puy_ytslice_256_512a_Om02}\end{figure}

Even though the value of $\Omega$ is rather high,
visualizations of the velocity near the surface only marginally
show signs of shear and/or rotation; see \Fig{pslice_256_512a_Om02}.
The plot shows that the convection is elongated in the east-west
direction (horizontal or $y$ axis).
Animations of the velocity on the periphery of the box
(see \url{http://www.nordita.dk/~brandenb/movies/conv-slab/chit})
as well as plots of $u_y(y)$ (see \Fig{pslice_lnrho_256_512a_Om02})
show that the random velocity dispersion ($\approx0.4\sqrt{gd}$) is still
large compared with the weaker shear flow of $\approx0.1\sqrt{gd}$
at the surface, and the even weaker prograde pattern speed
($\approx0.064\sqrt{gd}$).
The latter is best seen in space time diagrams of $u_y(y,t)$ for fixed
values of $x$ and $z$; see \Fig{puy_ytslice_256_512a_Om02}.
In this diagram one sees also several other slopes corresponding
to shorter-lived patters, some of which are even retrograde.
Similar values of the overall prograde pattern speed
have also been found for other latitudes, except that
the pattern itself is less clearly defined.
It is important to emphasize that the prograde pattern speed exceeds the
local shear flow at any depths inside the model, which is also what is
found for the sun.

%256_512a_Om02: VAR90

%256_512a_Om02: VAR90

\begin{figure}[t!]\begin{center}
\includegraphics[width=.7\columnwidth]{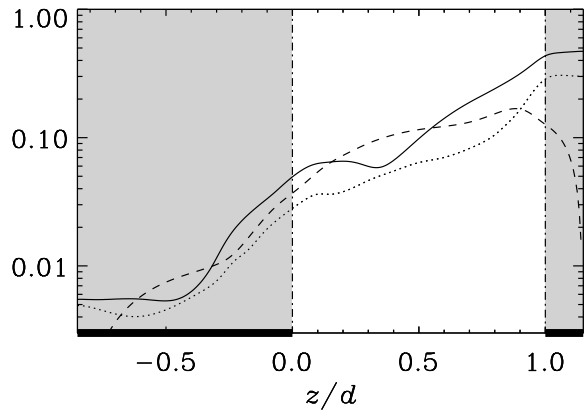}
\end{center}\caption[]{
Mean velocity dispersion (just for interest, not for the paper).
$0^\circ$ latitude.
$512\times512\times256$ resolution, $4\times4\times1$ aspect ratio.
}\label{ppuxyzrms_256_512a_Om02}\end{figure}

\begin{figure}[t!]\begin{center}
\includegraphics[width=.7\columnwidth]{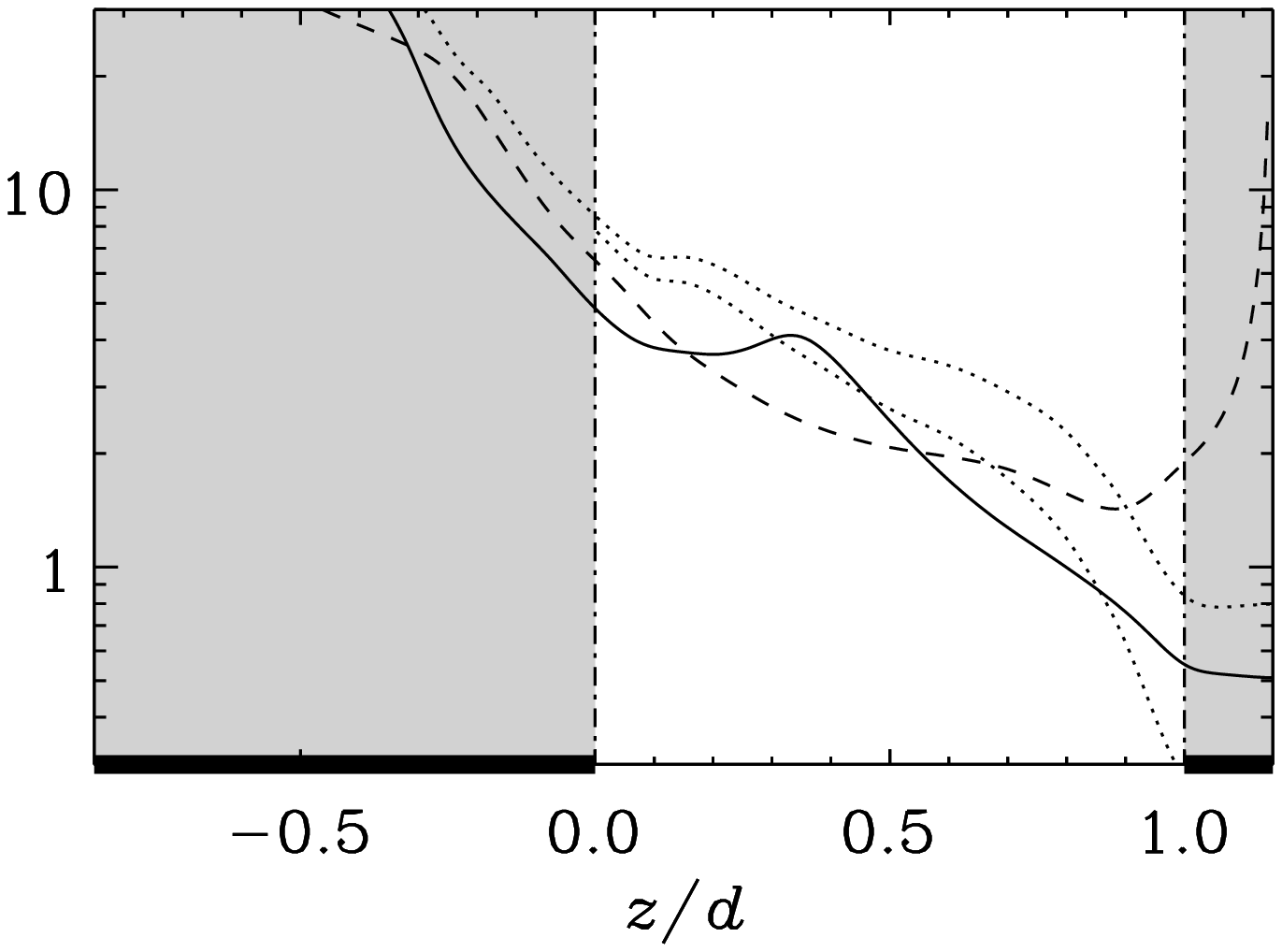}
\end{center}\caption[]{
Local Rossby number (just for interest, not for the paper).
$0^\circ$ latitude.
$512\times512\times256$ resolution, $4\times4\times1$ aspect ratio.
}\label{pprossby_256_512a_Om02}\end{figure}

\begin{figure}[t!]\begin{center}
\includegraphics[width=.7\columnwidth]{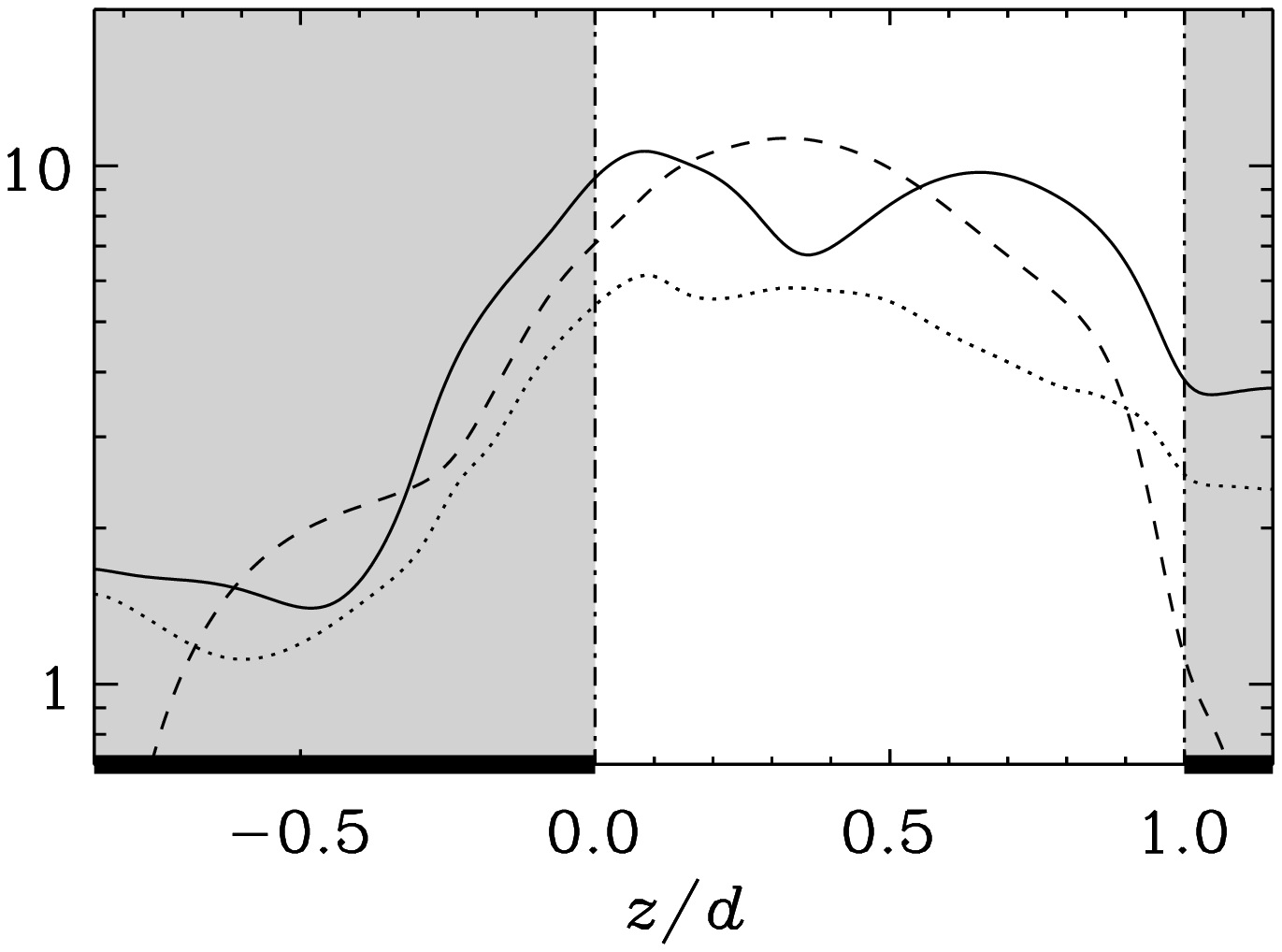}
\end{center}\caption[]{
Local Reynolds number (just for interest, not for the paper).
$0^\circ$ latitude.
$512\times512\times256$ resolution, $4\times4\times1$ aspect ratio.
}\label{ppuxyz_visc_256_512a_Om02}\end{figure}

Finally, we note that our simulations tend to show persistent vertical
p-mode oscillations with a frequency of about 2.5.
There is no indication that these modes are decaying in time,
although we need to remember that the resistive time, $L_z^2/(\nu\pi^2)$
($\approx1000$) is longer that the duration of the simulations.

\section{Implications for pattern speeds of magnetic tracers}

In addition to the recent suggestion that the supergranulation
may follow a prograde pattern speed, there are a number of
other indications pointing into the same direction.
Sunspots move faster than the gas at the surface, and younger
sunspots move faster than older ones.
This could be interpreted in two different ways.
Either the sunspot proper motion represents some real material
motion of the gas at some layer or, alternatively, it represents
some wave pattern speed, much like the patterns discussed here.
The latter seemed at first sight somewhat unexpected, given that
sunspots are generally believed to be rather more rigid and
coherent entities.
On the other hand, however, if sunspots represent actually more localized
phenomena, produced and sustained by local flow convergences,
it might appear plausible if sunspots and other magnetic features
are directly correlated with the prograde propagating pattern that
is seen in the simulations.

It is difficult to tell from simulations which picture is the
right one, unless one is able to obtain structures in turbulent dynamo
simulations that look like sunspots.
The idea of sunspots being a more localized phenomenon has been developed
by Kitchatinov \& Mazur (2000), who found that local sunspot-like flux
concentrations can develop near the surface as the result of a linear
instability in the thermo-hydromagnetic mean field equations.
Similar ideas have also been put forward by Kleeorin et al.\ (1996)
who drew a connections with a negative turbulent magnetic pressure
effect that might lead to sunspot-like flux concentrations.

The near-surface shear layer may be important for the solar dynamo in
producing toroidal magnetic fields provided appreciable mean poloidal
fields are present in the surface layer.
In fact, at the equator the amount of near-surface shear is at least
equally strong compared to that at the bottom of the convection zone, where at
$30^\circ$ latitude the shear is almost completely absent.
This and several other reasons give rise to the suggestion that the
solar dynamo may not just operate in the overshoot layer, but instead
in the full convection zone, and that a major fraction of the
toroidal field is being generated in the near-surface shear shear
(Brandenburg 2005).
This implies that an overshoot layer may not be a critical aspect of
the solar dynamo.
This is also consistent with recent findings by Dobler et al.\ (2006)
that even in fully convective spheres large scale magnetic fields is
produced.
Of course, before applying their model to real stars, it is important to
clarify the question of whether the dynamo still works at high magnetic
Reynolds numbers.

The original objection against dynamo action in the bulk of the
convection zone is connected with the possibility of a rapid loss of
magnetic flux through the surface via magnetic buoyancy.
The latter was never found to be a pronounced phenomenon in simulations
of compressible convective dynamo action, because stratified convection
leads to a strong downward pumping of magnetic fields
(Nordlund et al.\ 1992, Brandenburg et al.\ 1996).
This pumping phenomenon has recently also been confirmed in simulations
where an initial magnetic field was imposed, rather than generated by
dynamo action (Tobias et al.\ 1998, 2001, Dorch \& Nordlund 2001).
With this in mind, the idea of distributed dynamo action throughout the
entire convection zone has become viable again.

\section{Conclusions}

The present investigations have demonstrated that
in nonlinear rotating convection
(with gravity perpendicular to rotation, which is relevant to the equator),
negative shear emerges together with a positive pattern speed.
Although the mean azimuthal flow speed attains a positive maximum
away from the surface, its value ($\approx0.035\sqrt{gd}$ in
\Fig{ppuymzm_256_512a_Om02} at $z=-0.2d$) is still less than that of
the pattern speed ($\approx0.064\sqrt{gd}$).
Thus, the pattern cannot be associated with a particular depth where
the two would match.

Obviously, the usefulness of cartesian simulations for understanding
differential rotation is limited and comparison with global simulations will
remain necessary.
So far, global simulations have only shown traces of deceleration
toward the surface.
This difference could be explained by the absence of rapid downdrafts
near the surface, or perhaps by other differences in the setup of the
simulations (e.g.\ $\Omega$ is rather large in our simulations).
On the other hand, the basic aspects of the present simulations are also
seen in the more realistic solar simulations produced recently by
Stein \& Nordlund (unpublished) who find surface deceleration just like in
the present simulations.

\begin{discussion}

\discuss{J. Toomre}{There is a large body of linear stability studies,
in particular by Peter Gilman where he looks at the types of behavior of
various onset modes -- the equatorial modes and the high latitude modes
in a sphere.
It is characteristic of almost all of them that instead of them being
standing waves, they are propagating waves of convection.
They will be prograde at low latitudes and retrograde at high latitudes.
What is interesting is that in the big simulations some aspects of this
are still true.
The convection itself establishes strong differential rotation,
but the pattern itself also swims.
But on the issue that the supergranules are waves of something,
there is really some substantial debate -- the
supergranules can be swimming, but that does not make them waves.}

\discuss{A. Brandenburg}{Yes, I agree with you.
There is indeed a large body of analytical and linearized studies.
In my talk I mentioned also the recent papers by Busse (2004) and
Green \& Kosovichev (2006) that are certainly quite suggestive.
However, at the moment we cannot be sure that what we see in the simulations
is really just a nonlinear manifestation of the linear traveling wave patterns
seen in these analytical studies.
In the highly nonlinear regime the simulations show very sharp
structures that might resemble shocks.
So any direct relation to linear theory might be premature.
However, linear theory is the most obvious tool that we have at the moment.}

\discuss{M. Rempel}{How much faster is the fastest sunspot speed compared
to the mean flow.
When you have sunspots at low latitude, you also have this faster band of
the torsional oscillation.
I am wondering whether you might be able to reproduce the sunspot speed
if you add the additional velocity from the torsional oscillation to the
mean flow.}

\discuss{A. Brandenburg}{I think the torsional oscillations make a very
small effect here.
It is probably within the error bars seen in my plot.
The modulation with the cycle, which is what you are referring to,
is on the order of $\pm2\nHz$ at most.
This is small compared with the difference between the youngest and oldest
sunspots, which is of the order of $10\nHz$.}

\discuss{A.S. Brun}{You say that in order to break the constraint from
the Taylor-Proudman theorem you need are warm pole.
However, based on our various models we can say that it is
not just a question of having a warm pole, but it is a question of
having the right absolute temperature contrast ($\Delta T> 8\,$K or so).
In addition we have to consider the effects of the tachocline that may
have a feedback effect on the convection such as to give you the profile
we are observing in the sun.
So I do not think it is just a question of having a warm pole;
we can have a warm pole and still have almost cylinder-like rotation
if the baroclinic term is not strong enough.}

\discuss{A. Brandenburg}{My main point is that departures from
$\zzz\cdot\nab\Omega^2=0$ are most likely explained by a finite
baroclinic term, $\pphi\cdot(\nab T\times\nab S)$.
In the convection zone, where the radial entropy gradient is small,
a finite baroclinic term is mostly due to the latitudinal entropy
gradient, so that
\EQ
\varpi{\partial\Omega^2\over\partial z}\approx\pphi\cdot(\nab s\times\nab T)
\approx-{1\over r}{\partial s\over\partial \theta}{\partial T\over\partial r}<0,
\EN
where $\varpi=r\sin\theta$ is the cylindrical radius,
$z=r\cos\theta$ is the distance from the equatorial plane,
and $\pphi$ is the unit vector in the azimuthal direction.
Negative values of $\partial\Omega^2/\partial z$, in turn,
require that the pole is slightly warmer than the equator
(so weak that it cannot at present be observed).
Achieving this in a simulation may require particular care in the
treatment of the outer boundary condition.
I am aware that this is not really the case in the simulations, but
this would be the avenue along which one would hope to find an explanation.}

\discuss{A.S. Brun}{Your plot, where you show the rotation speeds of
new and old flux, seems to contradict your statement about the proper
motion of young and old sunspots.}
`
\discuss{A. Brandenburg}{These are really two different things.
The speeds of new and old flux refer to the speeds of active nests
that have life times of up to 6 months.
New flux appears at high latitudes where the angular velocity is less
than near the equator, where flux appears near the end of the cycle,
which is what is referred to as old flux.
Regarding sunspots, we are talking about much shorter time scales.
Young sunspots are those that have lived for 0.5 to 1.5 days, while
very old sunspots can live for 3 months.
The sunspot proper motion still obeys a latitudinal dependence very similar
to the average angular velocity, but the distribution of young spots is just
shifted upwards on such a plot.}

\end{discussion}

\end{document}